\documentclass[
reprint,
 amsmath,amssymb,
 aps,
]{revtex4-2}
\usepackage{graphicx}% Include figure files
\usepackage{dcolumn}% Align table columns on decimal point
\usepackage{bm}% bold math
\usepackage[english]{babel}

\usepackage{natbib}

\usepackage{datetime2}

\usepackage{letltxmacro}

\LetLtxMacro{\ORIGselectlanguage}{\selectlanguage}
\makeatletter
\DeclareRobustCommand{\selectlanguage}[1]{%
  \@ifundefined{alias@\string#1}
    {\ORIGselectlanguage{#1}}
    {\begingroup\edef\x{\endgroup
       \noexpand\ORIGselectlanguage{\@nameuse{alias@#1}}}\x}%
}
\newcommand{\definelanguagealias}[2]{%
  \@namedef{alias@#1}{#2}%
}
\makeatother

\definelanguagealias{en}{english}
\usepackage{amssymb}

\usepackage{float}

\usepackage{comment}

\usepackage{xcolor}

\usepackage{graphicx}% Include figure files
\usepackage{dcolumn}% Align table columns on decimal point
\usepackage{bm}% bold math
\usepackage[utf8]{inputenc}
\usepackage[T1]{fontenc}
\usepackage{mathptmx}
\usepackage{etoolbox}

\makeatletter
\def\@email#1#2{%
 \endgroup
 \patchcmd{\titleblock@produce}
  {\frontmatter@RRAPformat}
  {\frontmatter@RRAPformat{\produce@RRAP{*#1\href{mailto:#2}{#2}}}\frontmatter@RRAPformat}
  {}{}
}%
\makeatother

\begin{document}

\title{A Thermally Modulated SINIS Transconductance Amplifier}
% Force line breaks with \\
\author{G. Trupiano}
\email[Corresponding author: ]{giacomo.trupiano@sns.it}
 \affiliation{NEST, Istituto Nanoscienze-CNR and Scuola Normale Superiore, Piazza S. Silvestro 12, I-56127 Pisa, Italy}
 
\author{G. De Simoni}%
\affiliation{NEST, Istituto Nanoscienze-CNR and Scuola Normale Superiore, Piazza S. Silvestro 12, I-56127 Pisa, Italy}%

\author{F. Giazotto}
    \affiliation{NEST, Istituto Nanoscienze-CNR and Scuola Normale Superiore, Piazza S. Silvestro 12, I-56127 Pisa, Italy}

%\date{\today}

\begin{abstract}

We introduce a superconducting transconductance amplifier based on the thermal modulation of a SINIS (Superconductor–Insulator–Normal metal–Insulator–Superconductor) configuration. The device is composed of a normal metal island interfaced with two superconducting leads through tunnel barriers, establishing a voltage-biased symmetric SINIS setup. An additional NIS junction connects the island to a third superconducting lead, which serves as input. When the input voltage surpasses the superconducting gap, the resultant injection of quasiparticles increases the electronic temperature of the island, thereby modulating the SINIS current. We perform numerical analyzes of the device performance, influenced by input voltage, frequency, and bath temperature. At bath temperatures below $250~\text{mK}$, the device shows a transconductance exceeding $4~\text{mS}$ and a current gain exceeding $45~\text{dB}$. Both gain and transconductance maintain their levels up to $1~\text{MHz}$, but decrease at higher frequencies, with a $-3~\text{dB}$ cutoff around $10~\text{MHz}$, and an average power dissipation of approximately 5 nW. Our simulations reveal a fully voltage-controlled, three-terminal superconducting amplifier characterized by high transconductance and gain, achieved through thermally mediated signal transduction. This architectural design presents a promising avenue for cryogenic amplification with reduced power dissipation and compatibility with current superconducting electronic systems.
\end{abstract}


\maketitle

\section{Introduction} \label{sec:introduction}

Superconducting electronics has been identified as an emerging platform for quantum information processing and low-temperature sensing applications, due to its power efficiency, noise performance, and compatibility with quantum circuits \cite{doi:10.1126/science.1231930, golubevNonequilibriumTheoryHotelectron2001, SNSPD_Goltsman, giazotto2008ultrasensitive, korzhDemonstrationSub3Ps2020b}. A pivotal aspect of these systems is the amplification of weak signals while minimizing additional noise, particularly at millikelvin temperatures, where conventional semiconductor amplifiers exhibit suboptimal performance or require excessive cooling power as a result of their power dissipation \cite{paqueletwuetzMultiplexedQuantumTransport2020}. Although superconducting quantum interference devices (SQUIDs) have historically been the predominant choice for low-temperature amplification \cite{clarke2006squid}, their operation typically requires magnetic flux biasing, which complicates their integration into circuits. 

Various methodologies have been established for superconducting amplification without magnetic flux control, including Josephson parametric amplifiers (JPA) \cite{doi:10.1126/science.aaa8525, butseraenGatetunableGrapheneJosephson2022} and quantum phase-slip devices \cite{astafievCoherentQuantumPhase2012}. Although these devices frequently exhibit commendable noise performance, they have limitations in terms of bandwidth, dynamic range, or fabrication complexity. Field-effect devices, such as the Josephson Field-Effect Transistor (JoFET), utilize electrostatic control of superconductivity in low-dimensional materials or proximity-coupled weak links. This offers an alternative method for three-terminal superconducting devices, yet they encounter difficulties analogous to those faced by semiconductor electronics, particularly regarding dynamic power dissipation on the gate capacitance.

The modulation of temperature in superconducting devices \cite{morpurgo1998hot,roddaro2011hot} provides an alternative  mechanism for signal amplification. Within this category of devices, the introduction of charge or energy alters the electronic temperature or the distribution of quasiparticles in a superconducting component, thus influencing its transport characteristics \cite{1062499, booth_quatratran_2000, LEE2001981, seung-beckleeSuperconductingNanotransistorBased2003,tirelli2008manipulation,quaranta2011cooling, mccaughan_superconducting-nanowire_2014, baghdadi_multilayered_2020, buzzi_nanocryotron_2023, PhysRevApplied.23.014046}. Thermally operated devices possess distinct advantages, such as pronounced nonlinear responses and high input impedance, making them appealing for low-dissipation cryogenic electronic applications.

Normal metal-Insulator-Superconductor (NIS) junctions demonstrate a pronounced temperature dependence in their current-voltage characteristics, attributable to thermally activated tunneling of quasiparticles across the superconducting gap \cite{lemziakovApplicationsSuperconductorNormal2024a}. Superconductor-Insulator-Normal metal-Insulator-Superconductor (SINIS) structures have undergone a comprehensive investigation for their utility in electronic cooling applications, sensing devices, and thermometry \cite{giazotto_opportunities_2006}. The nonlinear temperature dependence observed in their current-voltage characteristics indicates that they could form the foundation for high-gain amplifiers with entirely thermal-controlled functionality. However, their viability as active electronic components remains significantly underexplored.

In this study, we introduce a superconducting transconductance amplifier engineered through the thermal modulation of a Superconductor-Insulator-Normal-Insulator-Superconductor (SINIS) configuration via quasiparticle injection through a third tunnel barrier on the normal metal island. This third barrier acts as the input electrode and ensures high subgap resistance, thus enabling effective thermal gating at low input power levels. This design also improves isolation and enhances compatibility with high-impedance circuits. Moreover, the operating voltage is approximately equivalent to the superconducting energy gap, which is substantially lower than that required for semiconductor transistors and Josephson field-effect transistors, thereby diminishing dynamic power dissipation.

We perform simulations of the amplifier static and dynamic characteristics, evaluating its performance in relation to input voltage, frequency, and bath temperature. Our findings underscore the promise of thermally gated superconducting devices for low-power, high-sensitivity amplification in cryogenic environments, particularly for applications in quantum information processing and ultrasensitive detection.

\begin{figure*}[t!]
\includegraphics[width=\linewidth]{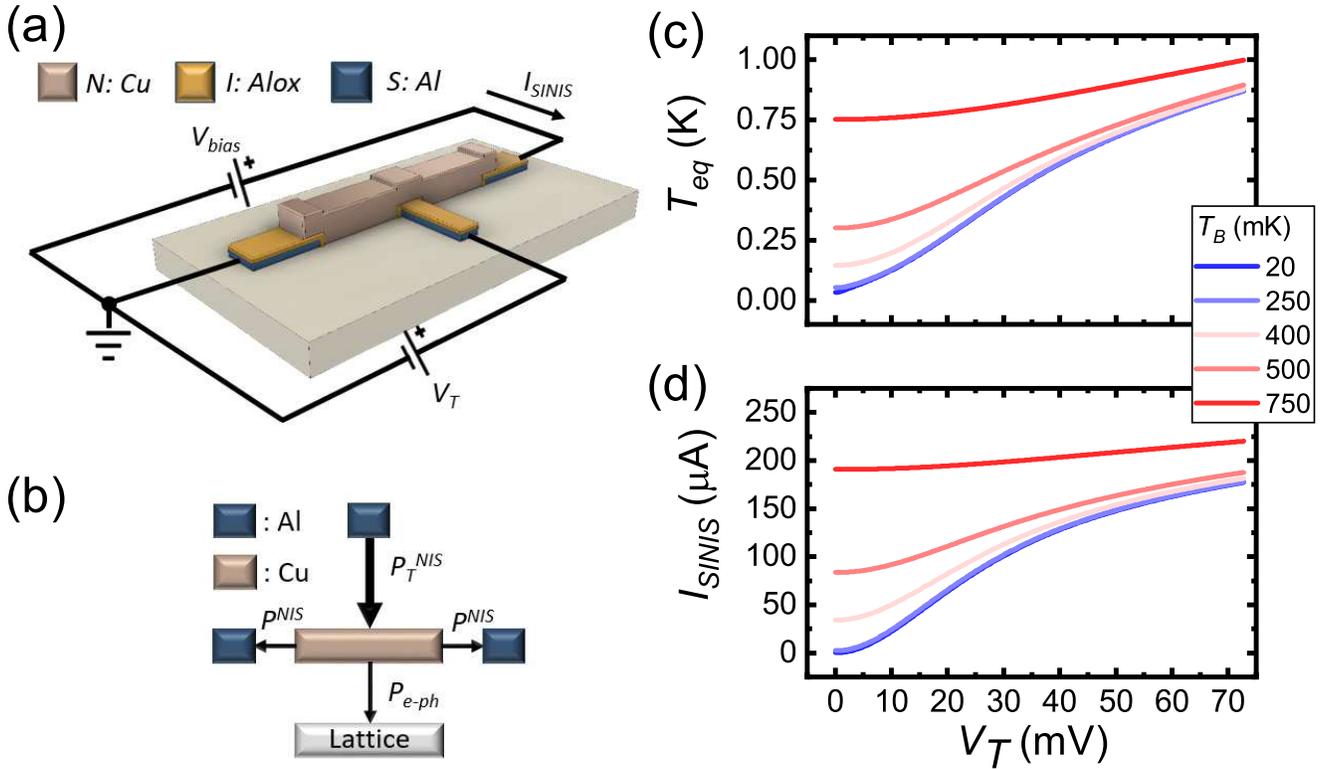}
\caption{\label{fig:fig1} Device scheme and working principle. (a) Device scheme: The device consists of two normal metal–insulator–superconductor (NIS) tunnel junctions connected in series, forming a symmetric SINIS structure. A normal metal (N, copper) island is located between two superconducting (S, aluminum) leads, with each interface separated by a thin insulating tunnel barrier (I, aluminum oxide). The SINIS is biased with a voltage $V_{\text{bias}}$, generating a current $I_{\text{SINIS}}$. The N island is also connected to a third superconducting lead via an insulating barrier, forming an additional NIS junction. This input junction modulates $I_{\text{SINIS}}$: when the voltage applied on the input junction $V_T/e > \Delta = 200~\mu \text{eV}$, quasiparticle injection heats the N island, affecting the SINIS current, which depends on the island electronic temperature.
(b) Heat current diagram: the equilibrium temperature of the N island ($T_{\text{eq}}$) is set by the balance of heat currents due to quasiparticle injection from the input junction ($P_T^{\text{NIS}}$), heat currents through the SINIS junctions ($P^{\text{NIS}}$), and electron-phonon coupling to the substrate ($P_{\text{e-ph}}$).
(c) $T_{\text{eq}}$ vs $V_T$ at various bath temperatures ($T_B$): $T_{\text{eq}}$ increases with $V_T$ due to quasiparticle heating. At higher bath temperatures ($T_B$), $P^{\text{NIS}}$ and $P_{\text{e-ph}}$ increase, reducing the heating efficiency of $P_T^{\text{NIS}}$. The bias voltage $V_{\text{bias}}$ was chosen to maximize the cooling power of the SINIS structure, with optimal cooling occurring at $V_{\text{bias}}\simeq 1.9\Delta(T_{B})$.
(d) $I_{\text{SINIS}}$ vs $V_T$ at various $T_B$: as $T_{\text{eq}}$ rises, the SINIS current increases.}
\end{figure*}
\section{Device Model} \label{sec:device model}

The device functions as a superconducting transconductance amplifier through thermal modulation of a SINIS structure. As depicted in Figure~\ref{fig:fig1}(a), the circuit schematic comprises two NIS tunnel junctions (normal metal-insulator-superconductor) arranged in series, thereby forming a symmetric SINIS configuration. This configuration includes a central normal metal island, composed of copper, which is connected to two superconducting leads, made of aluminum. The structure is biased by a voltage $V_{\text{bias}}$, enabling a current $I_{\text{SINIS}}$ to propagate through the device. A third superconducting lead is also tunnel-coupled to the N island via an extra NIS junction, which serves as the amplifier input. Due to the application of a voltage $V_T$, larger than superconducting gap ($V_T/e>\Delta$), across this input junction quasiparticles are introduced into the N island, thereby depositing energy and increasing its electronic temperature. The selection of aluminum and copper for our device arises because of their prevalent application in sub-kelvin superconducting electronics. Aluminum is particularly advantageous for the fabrication of high-quality tunnel junctions, offering reproducible characteristics \cite{giazotto_opportunities_2006}. Furthermore, the thermal properties of both materials at millikelvin temperatures are well documented and have been extensively characterized \cite{giazotto_opportunities_2006}. 

The steady-state temperature of the island $T_{eq}$ is established by balancing three thermal currents: the power injected from the input junction ($P_T^{\text{NIS}}$), the cooling power attributable to the SINIS structure ($2P^{\text{NIS}}$), and the thermal flow resulting from electron-phonon interactions coupling with the substrate ($P_{\text{e-ph}}$). Figure~\ref{fig:fig1}(b) presents this equilibrium through a simplified diagram depicting the pertinent thermal currents.
In our simulations, it was assumed that the electronic temperature of the superconducting banks is equivalent to the bath temperature, thereby effectively modeling the banks as infinite thermal reservoirs. Furthermore, it was assumed that the lattice phonons within the island maintain thermal equilibrium with the substrate at $T_B$. This assumption is substantiated by the minimal Kapitza resistance between thin metallic films and the substrate at sub-kelvin temperatures \cite{giazotto_opportunities_2006}.

The injection heat current flowing into the normal metal island is expressed by \cite{giazotto_opportunities_2006}:
\begin{align}
    P_T^{NIS} &= \frac{1}{e^2 R_T}\int_{-\infty}^{\infty}d\epsilon\,\tilde{\epsilon}\,\mathcal{N}_{S}(\epsilon-eV_T, T_B) \nonumber \\ 
    &\quad \times [f_0(\epsilon - e V_T, T_B) - f_0(\tilde{\epsilon}, T)],
\end{align}
where $e$ is the elementary charge, $R_T$ is the input tunnel junction resistance, $V_T$ is the voltage applied to the junction, $T$ is the electronic temperature of the island, $T_B$ is the bath temperature, $f_0(E, T)=1/[\exp{(E/k_B T)}+1]$ is the Fermi-Dirac distribution of the quasiparticles at energy $E$ and electronic temperature $T$, $k_B$ is the Boltzmann constant, $\mathcal{N}_{S}(E, T_B) = \left|\Re{\left[E+i\Gamma/\sqrt{(E+i\Gamma)^2 - \Delta^2(T_B)}\right]}\right|$ is the BCS density of states of the superconductor (smeared by the Dynes parameter $\Gamma=10^{-4}\Delta$) normalized at the Fermi level \cite{pekola2004limitations}, $\Tilde{\epsilon} = \epsilon - V_{bias}(R_{NIS}/R)$, and $V_{bias}(R_{NIS}/R)$ is the voltage drop across each tunnel junction. The latter depends on the ratio between the junction resistance $R_{NIS}$ and the total series resistance of the two junctions and the N island resistance ($R_N$), $R=R_{N}+2R_{NIS}$.

The dependence of the energy gap on electronic temperature was determined through a numerical approximation of the gap equation solution \cite{gross_anomalous_1986}:
\begin{align}
\Delta(T_B) \simeq \Delta_0 \tanh{\left(1.74\sqrt{\frac{T_c}{T_B} - 1}\right)},
\end{align}
where $T_c=1.3$ K is the critical temperature of an aluminum thin film of about 20 nm of thickness \cite{PhysRevB.26.3648}, and $\Delta_0\simeq 1.764 k_B T_c\simeq 0.2$ meV is the zero-temperature BCS gap. The heat current flowing through the SINIS structure depends on the electronic temperature of the N island and on the voltage applied across the structure, in particular on the voltage drop across the two tunnel junctions \cite{giazotto_opportunities_2006, giazotto_josephson_2005, savin_cold_2004}:
\begin{align}
    P^{SINIS} &= 2P^{NIS} = -\frac{2}{e^2 R_{NIS}} \int_{-\infty}^{\infty} d\epsilon\, \tilde{\epsilon}\, \mathcal{N}_{S}(\epsilon, T_B) \nonumber \\
    &\quad \times [f_0(\tilde{\epsilon}, T) - f_0(\epsilon, T_B)].
\end{align}
%where $\tilde{\epsilon} = \epsilon - V_{bias}(R_{NIS}/R)$, and $V_{bias}(R_{NIS}/R)$ is the voltage drop across each tunnel junction, which depends on the ratio between the junction resistance $R_{NIS}$ and the total series resistance of the two junctions and the N island resistance ($R_N$), $R=R_{N}+2R_{NIS}$.

The tunnel resistance values used in the simulations are denoted as $R_{\text{NIS}} = 0.5~\Omega$ and $R_T = 50~\text{k}\Omega$. In the proposed device fabrication procedure, we anticipate the implementation of two distinct oxidation processes: the first pertains to the SINIS structure junctions, characterized by a specific resistance of $30~\Omega/\mu\text{m}^2$, while the second involves a more intensive oxidation for the input tunnel junction, possessing a specific resistance of $2~\text{k}\Omega/\mu\text{m}^2$. Based on these postulations, the dimensions of the injection junction are represented by $200 \times 200~\text{nm}^2$. In contrast, those that pertain to the two junctions that comprise the SINIS structure are $6 \times 10~\mu\text{m}^2$. These tunnel resistance values and dimensions are compatible with existing devices as documented in the literature \cite{hatinenEfficientElectronicCooling2024a, giazotto_opportunities_2006}.

The normal metal resistance of the island is calculated as $R_N = \rho_{\text{Cu}}(l/wt) = 45~\text{m}\Omega$, where $\rho_{\text{Cu}} = 18~\text{n}\Omega\text{m}$ denotes the bulk resistivity of copper at 300 K, $l = 0.5~\mu\text{m}$ represents the length of the island, $w = 1~\mu\text{m}$ its width and $t = 0.2~\mu\text{m}$ its thickness. Using the bulk value for $\rho_{\text{Cu}}$ is a reasonable approximation for copper films with thicknesses exceeding 200 nm \cite{schmiedlElectricalResistivityUltraThin2008}. However, this value probably overestimates the actual resistivity at cryogenic temperatures, which is influenced by the residual resistivity ratio (RRR) of the material \cite{Ventura2014}.

The selection of the aforementioned resistance values is due to their significant impact on device performance. Specifically, a large transconductance ratio $R_T / R$ is advantageous, as high input resistance constitutes a fundamental attribute of an effective transconductance amplifier. Currently, the cooling power of the SINIS structure is essential in dictating the thermal relaxation dynamics of the normal metal island, which represents the main bottleneck that restricts the maximum operational speed of the device. Furthermore, it is imperative to maintain $R_{\text{NIS}} / R_N \gg 1$. This criterion ensures that most voltage drop occurs across the NIS junctions rather than the normal island, thus $V_{bias}(R_{NIS}/R)\simeq V_{bias}/2$. Given the non-linear dependency of the cooling power of an NIS junction on the voltage drop across each junction, reducing the voltage drop along the island enhances the amplifier overall performance. Moreover, the bias voltage $V_{\text{bias}}$ has been selected to optimize the cooling power of the SINIS structure, achieving the maximum effect at $V_{\text{bias}}\simeq 1.9\Delta(T_{B})/e$, thus further increasing the device operational speed.

The thermal current flowing out of the N island as a result of electron-phonon coupling is given by \cite{giazotto_opportunities_2006}: 
\begin{align}
    P_{e-ph} &= \Sigma V \left(T^5 - T^5_B\right).
\end{align} 
In this expression, $\Sigma$ represents the electron-phonon coupling constant, which for copper at sub-Kelvin temperatures is $\Sigma \sim 2\times10^9~\text{W}/\text{m}^3\text{K}^5$~\cite{giazotto_opportunities_2006,meschke_electron_2004}, and $V$ denotes the volume of the normal metal island, encompassing the copper regions that constitute the two NIS junctions within the SINIS configuration. The high electron-phonon coupling constant of copper contributes to enhance the amplifier bandwidth, since the ultimate limitation on operational frequency is attributable to the cooling time of the N island, as will be elucidated in the following.

As depicted in Figure~\ref{fig:fig1}(c), the equilibrium electronic temperature $T_{\text{eq}}$ increases with $V_T$ once $V_T/e$ surpasses the gap energy. The SINIS current $I_{\text{SINIS}}$ shows a pronounced temperature dependence, as the tunneling conductance of each NIS junction is nonlinearly modulated by the electronic temperature of the island. Consequently, controlling via $V_T$ the power of the injected quasiparticles modulates the SINIS current accordingly, culminating in a thermal transconductance amplification mechanism. This phenomenon is illustrated in Figure~\ref{fig:fig1}(d), where $I_{\text{SINIS}}$ is represented as a function of $V_T$. At low bath temperatures ($T_B$), heat leakage is minimal, and even a minimal $V_T$ can substantially increase $T_{\text{eq}}$, thus inducing significant modulation of the output current. As $T_B$ rises, the competing cooling powers $P^{\text{NIS}}$ and $P_{\text{e-ph}}$ increase, diminishing the net heating of the input and consequently the amplifier performance. This thermal gating principle facilitates a fully voltage-controlled, three-terminal superconducting amplifier that manifests high transconductance and current gain at cryogenic temperatures.
\section{Results and Discussion} \label{sec:results}

\begin{figure}[t!]
\includegraphics[width=\linewidth]{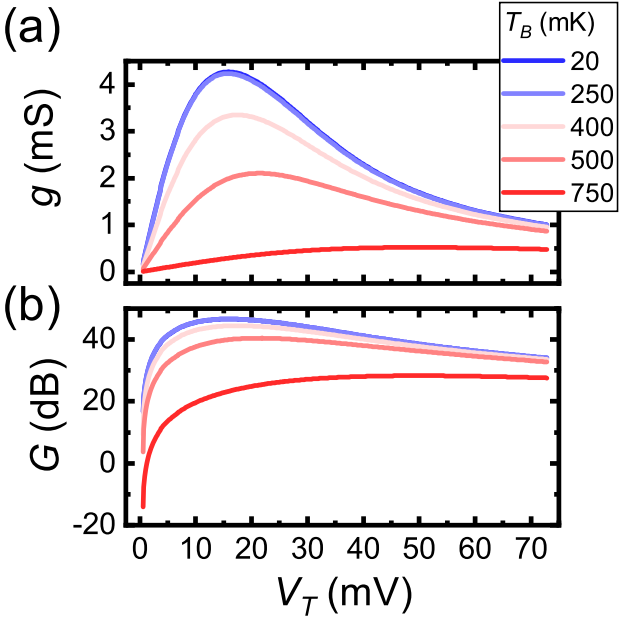}
\caption{\label{fig:fig2} Figures of merit: Transconductance and current gain as a function of $V_T$. (a) Transconductance ($g$) as a function of $V_T$ at different bath temperatures ($T_B$): for $T_B < 250$ mK, the maximum transconductance remains constant and exceeds 4 mS. As $T_B$ increases, $g$ decreases due to less efficient quasiparticle heating in the normal metal island. (b) Current gain ($G$) vs $V_T$ at different $T_B$: for $T_B < 250$ mK, the gain remains constant and exceeds 45 dB. Like transconductance, the gain declines at higher $T_B$ due to less efficient quasiparticle heating in the normal metal island.}
\end{figure}
The thermally modulated superconducting transconductance amplifier response to the injection of quasiparticles via the input NIS junction was examined. Principal performance metrics, specifically transconductance, current gain, frequency response, slew rate, and average dissipated power, were evaluated as functions of input voltage $V_T$, input frequency $f$, and bath temperature $T_B$.
Figure~\ref{fig:fig2} presents the static characteristics of the amplifier. 
As illustrated in Figure~\ref{fig:fig2}(a), the transconductance \begin{align}
g = \partial I_{\text{SINIS}} / \partial V_T
\end{align} exhibits a pronounced dependence on both $V_T$ and $T_B$. At low temperatures ($T_B < 250$ mK), $g$ exceeds 4 mS. This level of transconductance is comparable to that observed in commercial CMOS devices. With an increase in $T_B$, the peak transconductance diminishes, attributed to the reduction of the thermal efficiency of the input NIS junction. At higher temperatures, both the increases of the thermal dissipation, via the SINIS junctions, and of the electron-phonon coupling limit the rise of the electronic temperature of the normal metal island. A similar trend is observed in the current gain, defined as
\begin{align}
G = 20 \log_{10}(I_{\text{SINIS}} / I_T),
\end{align}
and shown in Figure~\ref{fig:fig2}(b). The gain exceeds 45 dB at low bath temperatures and decreases with increasing $T_B$, following the same thermal dependence as the transconductance. Here,
\begin{align}
I_T &= \frac{1}{e R_T}\int_{-\infty}^{\infty}d\epsilon\,\mathcal{N}_{S}(\epsilon-eV_T, T_B) \nonumber \\
& \quad  \times[f_0(\epsilon - e V_T, T_B) - f_0(\tilde{\epsilon}, T)]
\end{align}
is the current flowing in the input tunnel junction.

\begin{figure}[t!]
\includegraphics[width=\linewidth]{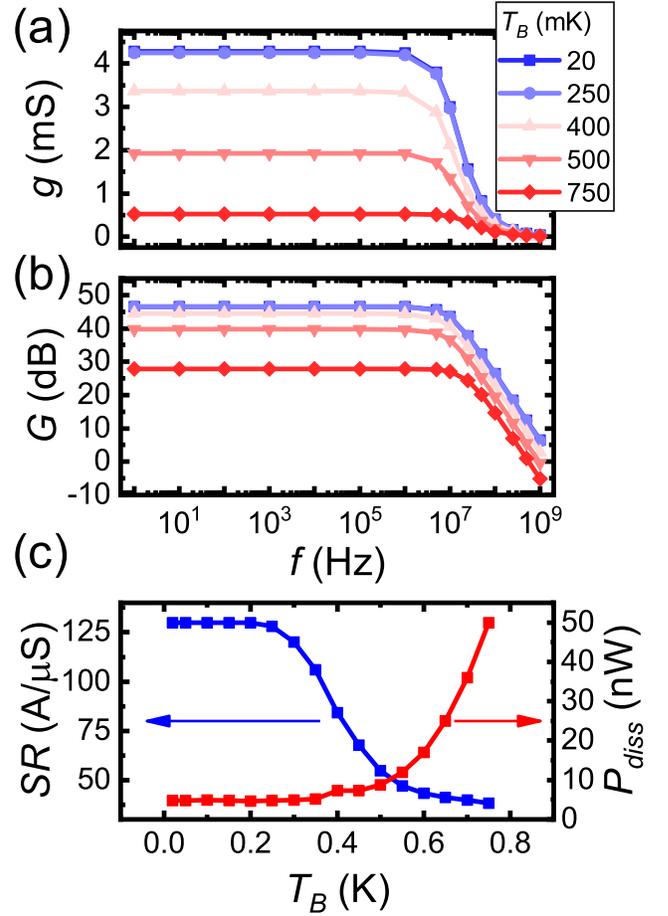}
\caption{\label{fig:fig3} Figures of merit: Transconductance and current gain versus frequency, slew rate ($SR$), and power dissipation ($P_{diss}$) against bath temperature.
(a) Transconductance ($g$) versus input signal frequency ($f$) at various bath temperatures ($T_B$): $g$ remains constant for $f < 1$ MHz and weakly depends on the bath temperature for $T_B < 250$ mK.
(b) Current gain ($G$) versus $f$ at different $T_B$: $G$ is constant up to $f < 1$ MHz and exhibits a cutoff frequency ($-3$ dB point) at approximately $f=10$ MHz.
(c) Slew rate and power dissipation versus $T_B$: The blue curve represents the slew rate, which stays roughly constant at 130 A/$\mu$s for $T_B < 250$ mK and then decreases monotonically as $T_B$ increases. The red curve indicates the average power dissipation, which is about 5 nW for $T_B < 250$ mK but rises at higher bath temperatures due to increased thermal leakage to the environment.}
\end{figure}

Figure~\ref{fig:fig3} shows the dynamic and temperature-dependent behavior of the amplifier. To sample these curves, we decided to use the peak of each gain and transconductance curve shown in Figure~\ref{fig:fig2} as the amplifier operation point $V_T^{DC}$. The amplifier current output as a function of a sinusoidal voltage input $V_T(t)=V_T^{DC}+A\sin(2\pi ft )$ was simulated by solving the heat equation of the N island:
\begin{align}
C_N (T) \frac{dT}{dt} &= P_T^{NIS}(T, V_T(t)) \nonumber \\
    &\quad - P^{SINIS}(T, V_{bias}) - P_{e-ph}(T),
\label{eq:Dynamics}
\end{align}
where the AC signal amplitude is set to $A = 0.5$ mV, and $C_N$ is the island heat capacity.
The heat capacitance was estimated using:
\begin{align}
    C_N(T)=\frac{\pi^2}{3}k_B^2 T V D(E_F),
\end{align}
where $k_B$ is the Boltzmann constant, $V$ is the island volume, and $D(E_F)=1.12\times 10^{47}$ 1/(Jm$^3$) is the density of states of copper at the Fermi energy ($E_F=7$ eV).

Figures~\ref{fig:fig3}(a) and~\ref{fig:fig3}(b) show that both $g$ and $G$ remain invariant up to approximately 1 MHz, suggesting that the thermal dynamics is capable of tracking the input voltage variations within this frequency range. Beyond the threshold of 1 MHz, the gain diminishes, exhibiting a $-3$ dB cutoff frequency around 10 MHz. This attenuation is attributed to the thermal response time of the normal metal island, mainly restricted by its cooling dynamics at sub-Kelvin temperatures~\cite{giazotto_opportunities_2006}. Furthermore, Figure~\ref{fig:fig3} illustrates the temperature dependence of $g$ and $G$. At $T_B < 250$ mK, both parameters display a weak dependence on bath temperature, whereas at higher $T_B$, they decrease monotonically due to increased thermal leakage.

%Finally, Figure~\ref{fig:fig3}(c) reports the temperature dependence of the slew rate ($SR$) and the rise time ($t_r$). The slew rate remains constant at low $T_B$ and decreases as the temperature increases, consistent with the drop in $g$ and the thermal sensitivity. The rise time shows a non-monotonic behavior: it remains stable at low $T_B$, peaks near 500 mK, and then decreases. This behavior likely reflects a complex interplay of increasing heat capacity, nonlinear thermal conductance of the SINIS junctions, and a reduced temperature gradient between the island and the bath at higher temperatures.

Ultimately, Fig.~\ref{fig:fig3}(c) illustrates the relationship between the slew rate ($SR$) and average power dissipation ($P_{diss}$) as functions of the bath temperature, while the amplifier operates at an input voltage corresponding to the maximum transconductance. The slew rate attains its peak value of approximately 130 A/$\mu$s at a bath temperature of $T_B < 250$ mK and subsequently diminishes as the bath temperature increases. 
The power dissipation $P_{diss}$ has been evaluated as the sum of all heat current contributions: 
\begin{align}
    P_{diss} = P_T^{NIS} + P^{SINIS} + P_{e-ph}.
\end{align}
Its magnitude is approximately 5 nW for bath temperatures below 250 mK. Nevertheless, it scales with increasing bath temperature due to enhanced thermal leakage, necessitating a higher input voltage to sustain the device in its optimal operational state. However, the power dissipation remains several orders of magnitude lower than that encountered in conventional cryo-CMOS amplifiers. 
In such systems, the primary factor in power consumption is the dynamic power dissipation \cite{paqueletwuetzMultiplexedQuantumTransport2020}, which can reach magnitudes of 1 nW/Hz, equating to 1 mW at a frequency of 1 MHz. In this device, the dynamic power dissipation is negligible, as it operates with voltages on the scale of 1 mV. The dynamic power can be calculated as $P_{dyn}=CV_{AC}^2f \simeq 2$ fF$\times ($1 mV$)^2\times$ 1 MHz$ = 2$ fW. Here, $C$ is the input electrode capacitance, estimated using the expression $C = c_s A_{in}$, where $c_s = 50$ fW/$\mu\text{m}^2$ is the specific capacitance of a typical aluminum oxide tunnel junction \cite{10.1063/1.2357915}, and $A_{in} = 0.2\times0.2$  $\mu$m$^2$ is the junction surface area. Hence, power dissipation is mostly unaffected by the operating frequency of the device.

The linearity of the amplifier was evaluated through the simulation of the total harmonic distortion (THD) of the output signal, defined as: 
\begin{align}
L_{THD} = -20\log{\left(\frac{\sqrt {\sum_{i=2}^{\infty} A_i^2}}{A_1}\right)}.
\end{align}
Within this context, $A_i$ represents the amplitudes of the Fourier components of the output signal, and $A_1$ denotes the amplitude of the fundamental frequency. The results yielded linearity values within the range of 50–60 dB. However, these figures should be regarded as upper limits. This consideration arises from the finite sampling rate employed in the simulation, which may lead to an underestimation of the amplitude of higher-order harmonics.
\section{Conclusion} \label{sec:conclusion}

In conclusion, we have introduced an innovative superconducting transconductance amplifier that operates on the principle of thermal modulation within a symmetric SINIS structure. The system is regulated through quasiparticle injection from an auxiliary terminal on the normal metal island. Using the thermal sensitivity of the central normal metal island and isolating the control input from the amplifier output through tunnel barriers, the device successfully achieves voltage-controlled amplification with minimal leakage and dissipation. Our simulations indicate a high differential transconductance reaching up to 4 mS, alongside a significant current gain of up to 45 dB at cryogenic temperatures, while maintaining stable performance within the megahertz frequency domain.

The operational parameters of the device are primarily dictated by the thermal dynamics of the island, wherein the regime of operation is characterized by the interaction between electron-phonon cooling, tunnel-mediated heat extraction, and heat injection by quasiparticles. Although the device exhibits temperature sensitivity, it demonstrates consistent performance up to a bath temperature of 250 mK, making it highly compatible for integration into cryogenic systems where conventional semiconductor amplifiers often prove inadequate. Furthermore, the amplifier maintains an average power dissipation below 5 nW, which is considerably lower than that of standard cryo-CMOS circuits and remains unaffected by variations in operating frequency.

This study establishes the groundwork for developing low-power, high-sensitivity amplifiers in superconducting nanostructures based on thermal gating. The fully voltage-controlled, three-terminal architecture enables integration with superconducting and quantum technologies. These attributes make our amplifier a promising foundational component for future cryogenic readout architectures and quantum information processing systems. This device is optimally designed to modulate current within a low-impedance load using a minimal input voltage. It is applicable for driving a superconducting flux line or for precisely adjusting the bias current of superconducting photon detectors, including transition edge sensors \cite{10.1063/1.113674, paolucci_development_2020} (TES) and superconducting nanowire single-photon detectors (SNSPDs) \cite{SNSPD_Goltsman, korzhDemonstrationSub3Ps2020b}. Furthermore, it is suitable for driving the input current of a nanocryotron \cite{mccaughan_superconducting-nanowire_2014}.

\section*{Acknowledgements} \label{sec:acknowledgements}

We acknowledge the EU’s Horizon 2020 Research
and Innovation Framework Programme under Grants No.
964398 (SUPERGATE), No. 101057977 (SPECTRUM),
and the PNRR MUR project PE0000023-NQSTI for partial
financial support.

\end{document}